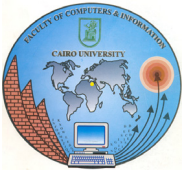
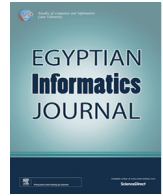

# Lane prediction optimization in VANET

Ghassan Samara

*Department of Computer Science, Zarqa University, Jordan*

ABSTRACT

Among the current advanced driver assistance systems, Vehicle-to-Vehicle (V2V) technology has great potential to increase Vehicular Ad Hoc Network (VANET) performance in terms of security, energy efficiency, and comfortable driving. In reality, vehicle drivers regularly change lanes depending on their assumptions regarding visual distances. However, many systems are not quite well-designed, because the visible range is limited, making it difficult to achieve such a task. V2V technology offers high potential for VANET to increase safety, energy efficiency, and driver convenience. Drivers can make more intelligent options in terms of lane selection using predicted information of downstream lane traffic, which is essential for obtaining mobility benefits. An assistant lane selection system is proposed in this research, which helps the driver locate an optimal lane-level travel path in order to minimize travel time. The decision-making criteria are based on the predicted lane traffic conditions via V2V technology. This paper aims to create a specific V2V system to support lane selection based on the predicted traffic states to find the best travel lane. In this paper, a Spatial–Temporal (ST) prototype is developed and then applied to predict future traffic conditions for road cells using spatial and temporal information. The suggested lane selection assistance system uses this information to select the optimized lane sequence. Then, an intensive simulation-based assessment is conducted in different scenarios. Results indicate that the proposed system outperforms other published systems.

© 2020 Production and hosting by Elsevier B.V. on behalf of Faculty of Computers and Artificial Intelligence, Cairo University. This is an open access article under the CC BY-NC-ND license (http://creativecommons.org/licenses/by-nc-nd/4.0/).

## 1. Introduction

In recent years, advanced driving aid schemes have been influenced by Connected Vehicle (CV) technology, which allows wireless communication between cars and between cars and facilities (i.e., Vehicle-to-Everything or V2X). To improve the security, effectiveness, and convenience of riding vehicles [1,2] different trials on V2X information return methodologies have been conducted [3,4]. Current CV apps are primarily intended for two kinds of communication: Dedicated Short-Range Communication (DSRC) and mobile-oriented communication depending on Wireless Access in the Vehicular Environment (WAVE) [5]. By using the IEEE 802.11p norm [6] the DSRC systems can provide elevated accessibility and low-latency streams for critical security apps [7]. However, they require comparatively costly embedded units for all communication-capable terminals [2]. Due to the accessibility of several embedded detectors, mobile devices (e.g., smartphones) can readily be incorporated with separate advanced driving aid schemes [8,9]. The CV technique has drawn enhanced interest because of its ability to improve car security and safety [10,11] economic sustainability, and riding comfort [1,12]. In [10] the authors concentrated on the safety of road lane change and built secure pathways for conductors using predictive power models. In order to properly direct the rider around the junction in an environmentally friendly manner, the authors of another work [13] suggested an eco-approach and departure request, which could obtain signaling stage and time data from the upcoming traffic signal. One study [14] suggested an in-car scheme that utilizes the position of traffic light and time to reach an optimum personal riding rate. In addition, various organizations have created several efforts to encourage CV studies. For example, a wide range of apps created under the safety pilot program [15] the dynamic mobility application program, have been described in the CV reference implementation architecture [16]. Examples of environmental applications include the real-time information synthesis [17] program and US funding program for road weather CVs [18] and the US Department of Transport (USDOT). Moreover, several studies on CV apps have been financed by the European Union and other countries [19].

The vehicular ad hoc network (VANET) faces several difficult situations, which differentiate it from other Mobile Ad Hoc Networks (MANETs). A highly complex network topology is accomplished by the amount of traffic under varying conditions, such as hours of traffic and traffic jams. Furthermore, the high mobility feature of the vehicles results in an intermittent communication between vehicles and/or between cars and Roadside Units (RSUs) [20,21]. Due to the behavioral instability of close-knit human drivers, changing lanes has become a critical bottleneck in the secure deployment of autonomous vehicles. Human drivers can adaptively respond to the increasingly differing traffic situations according to their understanding and experiences. However, it remains difficult to illustrate clear rules for all scenarios from gathered raw data, simply because the data flood can overpower human insight and interpretation. In this paper, a lane selection system is proposed, which helps drivers find an ideal lane route to minimize journey time. The decision-making process is focused on the prediction of lane-level traffic by means of CV technology.

The rest of this paper is organized as follows: Section II presents the related works, followed by problem formulation in Section III. Section IV presents the detailed description of the lane selection algorithm. In Section V, simulation studies are conducted to evaluate the performance of the proposed system. The final section concludes this paper with further discussion on future works.

## 2. Related Work

A wide range of CV apps for riding help have been suggested and created, but only a few of these have focused on side control support. Examples include the allocation of the zone [22,23] and the choice of an ideal zone [24] as suggested by past works [22,23] by means of inter-vehicle communications, which is a decentralized track strategy to a set of cars and car pilots on/off roads. An additional track choice study project has been suggested to regulate uncoordinated track modifications by means of two-way communication between vehicles and infrastructures (vehicle-to-infrastructure, V2I) by minimizing possible clashes in the car [23]. The findings in [23] indicated that the median journey moment is relatively lower with the non-lane choice situation as a result of controlled lane change behaviors. All of these studies assumed that all cars on the highway are applied vehicles, making it difficult, in the following decades or more, to achieve these vehicle allocation methods. Yet, to date, the movement has not been well researched in the area of advanced driving aid schemes featuring directional command help. Meanwhile, traffic-state predictions, such as sampling Kalman, non-parametric correlation model, or cellular networks, have been well-researched for years [25,26]. One study [25] proposed a linear regression model relying on results from loop detectors for transport moments in the freeway. Another work [27] suggested a straightforward, stable time series system for a segment of a motorway to predict transport times. Many model- and data-driven models, such as the hidden Markov models [28,29] K-nearest neighbors method to traffic-state forecast [30] the particle filter model [31,32] the Kalman filter [33] and deep neural networks in [34] have also been suggested for short-term flow government forecast.

A number of vehicle prediction techniques were created on the basis of the Markov chain [28,35]. The Markov (binary) system was used mostly for determining the vehicle state of the next interval depending on the signal models. In conjunction with the Markov variable-length strings, for instance, the closest neighboring ranking was used to forecast traffic patterns [28]. Following classification into a group of the vehicle status for each fresh time step, a specific speed score is calculated using the suitable weighted regression model tracked only with information from the corresponding group. In order to ensure the elevated predictability of the short-term traffic flow predicting technique [35] a mixed forecast technique centered on Markov's loop hypothesis and Grey Verhulst's model was also suggested. The volatility of the information was addressed in the Markov chain principle, which is based on the Gray Verhulst model, in an attempt to enhance the precision of the forecasts.

Findings showed that a relative error of traffic flows from one section over 16 times (5 min per time step) between 0 and 13 times (between real-world information and predictive information). All previous studies, however, relied on a Markov chain although it is not very helpful in describing events and, in the majority of cases, cannot be the actual model of the underlying situation. At the same moment, a short-term forecast technique depending on space–time correlations was also suggested. In [36] the authors pointed out that the transport status of a particular location has been extremely influenced by flow upstream and outgoing circumstances and that free flow rates (cell-to-cell, lane-to-lane correlation) are spatially linked. An extended stochastic cell transmission model has been used in support of short traffic state prediction in light of the spatial time correlation of the transmission traffic flux. In [36] the I210-W segment was split into four cells with a per-cell range of around 0.5 miles. In order to validate the efficacy, the general average relative percentage error has been calculated, ranging from 10.8 to 14 times [36]. In [37] an assessment method for traffic statuses using correlations between highway networks and scarce traffic samples was suggested in order to assess traffic circumstances in various highway sections. Their suggested method focused on the mathematical structure of the multi-linear regression (MLR) representing road connections. The MLR system and the compressive detection technology were used to estimate congestion at urban scale by monitoring a tiny amount of sampled cars. A comprehensive experimentation on real-world traffic information (within a big network of 1826 highway sections in the town of Shanghai) helped validate the template for traffic estimations. Results showed that, for various road situations, the complete velocity difference between the findings expected and the soil truths is 5.2–11.0 km/h. However, the experiment should be further investigated for higher speeds to further test the results.

Another way of estimating and predicting the road state is to use an enhanced Kalman Filter ensemble to assess and forecast realistic highway network which, thanks to lower matrix reversals, can decrease computing time [32]. However, Kalman Filter considers both equations between the device and the monitoring model to be linear, which in many real-life circumstances, is not realistic. In [34] the development of the road state in a highway was modelized through a computer training neural network. All these methods are concentrated instead of lane point on the forecast of link-level congestion state.

In recent years, lane-based surveys have drawn increasing interest, including, but not limited to projections of car travel, queue alert efficiency assessment, and the lateral movement prediction of independent cars [38,39]. Lane-level guide techniques, such as those using improved global positioning system (GPS)/multilayer chart systems for ego-lane evaluation and lane-level mapping system, have also been proposed to facilitate the execution of lane-level apps in practice [40,41]. However, GPS does not scale up with high speed networks like VANET. Advanced detecting systems, such as the radio-frequency identification (RFID) technology [42,43] have also been created as the main facilitator for precise location monitoring, which can help countless transit apps in the future [44]. The RFID technology, however, is slower compared to VANET. In [45] the OLS application has been proposed for achieving optimal lane selection. For practical usage, the utility function has been built with longitudinal and lateral safety considerations in mind. The OLS app was initially assessed in a highway segment. Results revealed that the OLS implementation reduced travel time and delay compared to the base case without control inside the linked transportation system. The application also improved efficiency on such higher traffic demands, but not under capacity conditions, which caused lane changes to prevent CVs from moving in congested areas. High-efficiency lane-level prediction can help automated CVs select and plan the optimum lane paths based on the predicted traffic flow in terms of the level of service. Sequentially, a more balanced overall distribution of different vehicles can be achieved and, correspondingly, an increased capacity for roads [46]. Inspired by this study, we suggested a model of regression in the forecast of lane-level state vehicles through the use of inter-lane data (intra-lane data) and inter-lane interconnections between neighboring highway sections on the same lane.

## 3. Problem formulation

In real conditions, riders generally change distances, many of which are not well-planned, depending on their assumptions. Consider, for instance, a decider car (the car of concern for individuals) riding a five-lane highway under high congestion circumstances (see Fig. 1). The driver's destination zone has to decide which spot to shift to (i.e., the first choice room in lane three or second choice room in lane five).

As the traffic upstream of the decider car (in lane 4) is congested in the driver's sight, it is difficult for the rider to precisely understand which lane has heavier vehicles because of the restricted viewing range. Assuming that the decider in lane four switches to room in the second choice and faces two options, either to continue in this blocked lane or go back to the previous lane, the rider is likely to switch back to his previous lane. The rider may also decide to switch to space first choice in lane three, as there is more room there. In this case, it may be a better option to stay in lane three or the decider may switch the room in lane 2. Therefore, anticipated upstream lane-level data are vital to achieve mobility advantages, which would allow deciders to create better decisions in route choice.

We describe as communications-capable those cars that can communicate fundamental data (i.e., speed and position). RSUs are small routers placed on the roads [21]; vehicles send beacon messages to RSUs periodically at a frequency of ten messages every second containing specific data (i.e., speed, position, and direction) [47]. Fig. 2 presents the beacon structure.

Note that road conditions change over a moment, so vibrant designs are necessary to predict traffic conditions. The forecast for the road level can be performed by using regression designs. The optimization problem is developed to determine what room (i.e., part of the range) the car should occupy at some stage in order to find the finest range route for an application.

See Fig. 3 for the proposed information flow.

## 4. Proposed system

### 4.1. Beacon upload

Each vehicle uploads ten beacons per second to an RSU; the received beacons by the RSU are then used to analyze driver pattern and current traffic conditions. The beacons received will be analyzed in two steps: (1) obtaining the average of the current speed (ACS) by computing the current average speed of the sender vehicle in relation to the average speed of other cars in the same location and (2) comparing that to the traffic density. This can be calculated using Eq. (1):

$$ACS = \frac{\sum \frac{S}{T}}{N} \quad (1)$$

where S is the vehicle speed, T is the time, and N is the number of current vehicles connected to the RSU. The ACS ratio is then compared to the average of all vehicles speed (AAVS) registered to the RSU. The AAVS can be computed using Eq. (2):

$$AAVS = \frac{\sum_1^n ACS}{N} \quad (2)$$

where AAVS computes the average speed of all vehicles connected to the current RSU in the network. Eq. (3) compares the ACS to AAVS:

$$Anlaysed\,speed = if(ACS > AAVS) \quad (3)$$

If Eq. (3) is fulfilled, then the current vehicle is faster than the average of the current vehicles on the network. The AAVS must take into account the percentage of sudden moves that could occur in the current situation, such as sudden breaks or sudden lane changes at high speeds performed by any driver. Using Eq. (4), we can calculate the overall average for a particular vehicle:

$$AvSud = \left(\sum_{k=0}^{n} Sud.brk/n\right) + \left(\sum_{k=0}^{n} Chg.Loc/n\right) \quad (4)$$

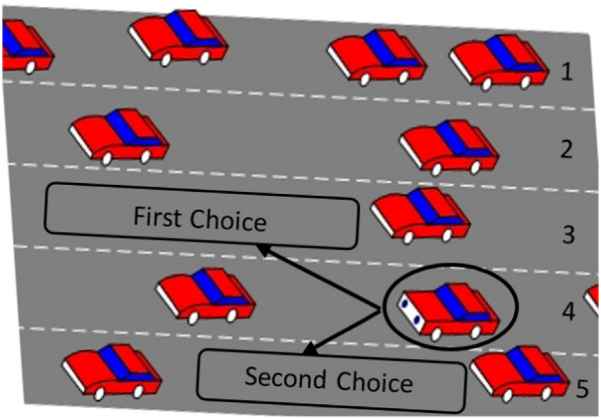

Fig. 1. An example of the problem description.

Fig. 2. Beacon structure.

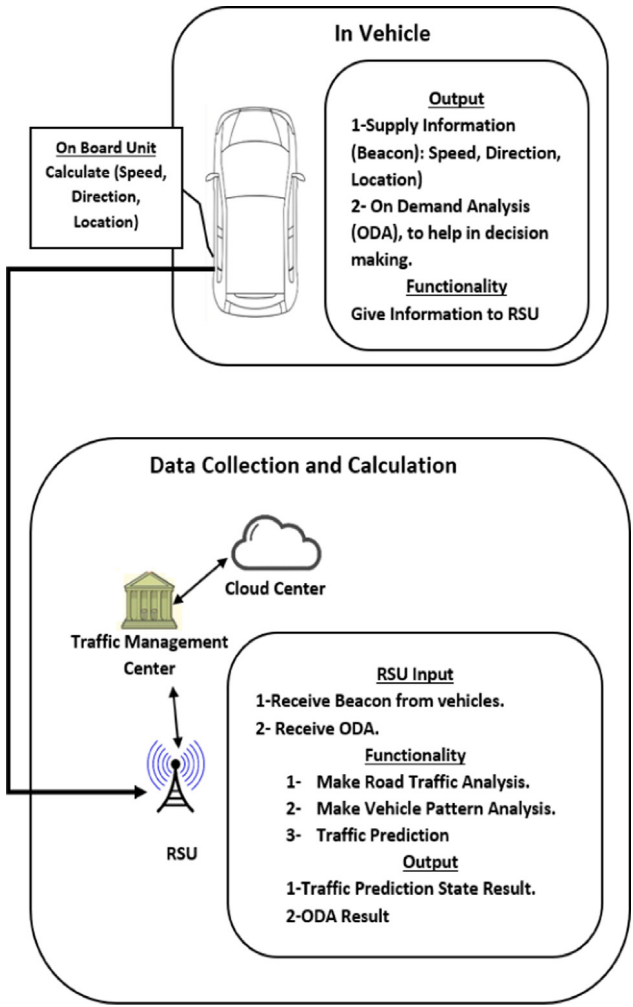

**Fig. 3.** Information flow of the lane selection.

where n is the number of times the sudden action is measured for each car, sud.brk is the number of sudden breaks performed by that driver at high speed, and Chg. Loc is the number of sudden lane changes executed by this driver at high speed. If the average AvSud in the current network is high, then the unexpected behavior can be predicted to suddenly change the shape of the traffic. Going back to the example in Fig. 1, the circled vehicle has two options for selecting the first or second choice: either the vehicle will send ODA to RSU or the ODA will contain beacon and ACS about decider vehicle. The RSU will then evaluate the vacant choices for each option by measuring the current AAVS and the current average AvSud. As the vehicles around each choice vary, this can result in some AAVS and AvSud reaching the results listed in Table 1.

In the first and second case, the decider is slower than the vehicle around him, and will not be given the opportunity to move to the vacant position as other vehicles may move before him. Therefore, this choice is not preferred.

**Table 1**
Traffic prediction decision.

|   | ACS > AAVS | AvSud | Decision |
|---|---|---|---|
| 1 | No  | High | Not preferred |
| 2 | No  | Low  | Not preferred |
| 3 | Yes | High | Not preferred, possible danger |
| 4 | Yes | Low  | preferred |

In the third case, the vehicle has a higher speed than other vehicles, but there is a high likelihood of sudden breaks or lane changes, so that there is a risk of possible crashes. Therefore, this choice is not preferred. For the fourth choice, the decider vehicle is faster than other network vehicles, and the other vehicles have no habit of changing lanes or unexpected breaks; in this case, this option is preferred. Hence, the RSU will send its analysis to the decision-makers' vehicle to help it decide which choice is better in terms of fulfilment and safety.

## 5. Experiments, results and analysis

Intensive simulation with the new MATLAB R2019a was performed to demonstrate the accuracy of the proposed protocol [48]. The new interactive wireless environment is included in the MATLAB R2019a, and the same environment has been created with the same parameters. The implementation focuses on showing a boost in the system's performance in relation to the proposed St-Model [46] framework. The simulation included 25,000 runs to achieve more precise results. The parameters of the simulation for the entire experiment are shown in Table 2.

In the first experiment shown in Fig. 4, we tested the relationship between congestion and travel time, which in high-traffic conditions, is expected to increase. There are three levels in the test of congestion (traffic density): low − 25% of the total number of vehicles in the experiment, medium − 50% of the total number of vehicles in the experiment, and high − 75% of the total number of vehicles in the experiment. The experiment shows that the St-Model has scores ranging between 9% and −7%, with a mean of −8% of the total journey time when the level is low. The proposed model's scores vary from −10% to −8%, with a mean of −9% of the total journey time. When the level is medium, the St-Model ranges from −7% to −3%, with a mean of −5% of the total journey time. Meanwhile, the proposed model varies from −9% to −5% with a mean of −7% of the total journey time. At a high level, the St-Model scores range between −3% and 3%, with a mean of −1% of the overall journey time. The proposed model's scores vary from −8% to −2%, with a mean of −5% of the total journey time.

This experiment demonstrates that the proposed system produces better results in heavy traffic situations, because it collects richer information on traffic conditions using ODA, which in turn, leads to better choices.

**Table 2**
Simulation parameters.

| Parameter | Value |
|---|---|
| Simulation Grid | 1000 × 1000 |
| Simulation time | 300 sec |
| Vehicle speed | 15–45 m/s |
| Number of runs | 25,000 veh/run |
| Maximum Number of vehicles | 200 |
| Number of lanes | 6 (3 in each direction) |
| Scenario | Two-way highway |
| Network interface | Phy/WirelessPhyExt |
| MAC interface | Mac/802 11Ext |
| Interface queue | Queue/DSRC |
| Propagation model | Propagation/Nakagami |
| Number of TDMA slots/frames | 10 |
| Time slot | 2.5 ms |
| Message size (safety) | 100 bytes |
| Message size (nonesafety) | 512bytes |
| Transmission range | 300 m, 500 m |
| Modulation type | BPSK |
| Antenna type | Antenna/omniantenna |
| Channel type | Channel/wireless channel |
| Data transfer rate | 6, 12, 18, 27Mbps |
| Minimum beaconing interval | 100 ms |
| Maximum beaconing interval | 500 ms |

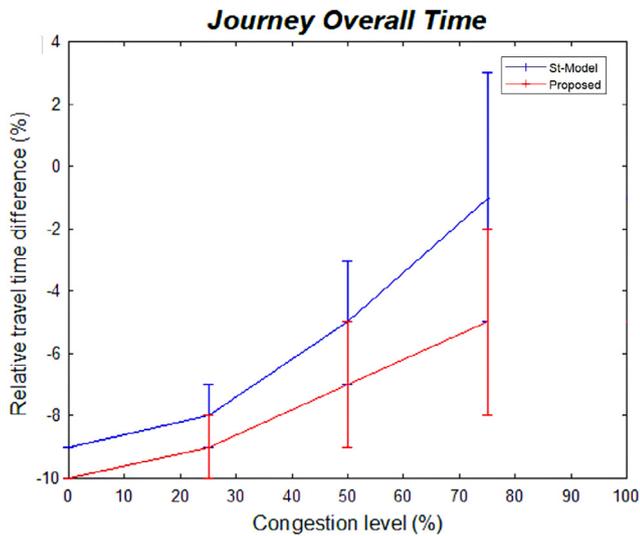

Fig. 4. Congestion effect on travel time.

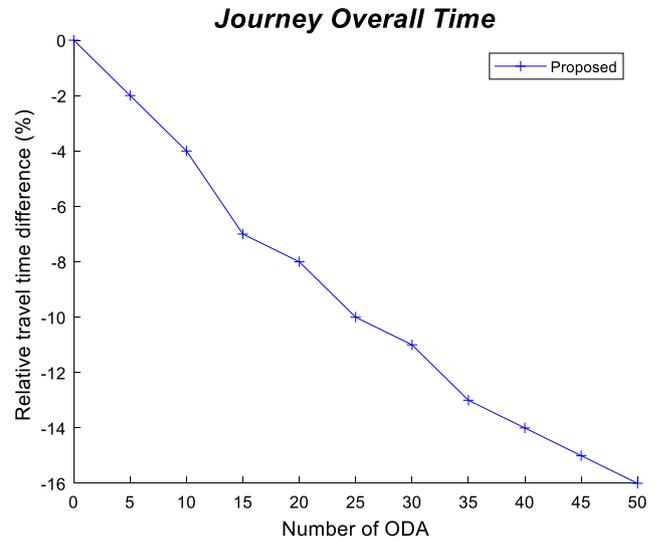

Fig. 6. Impact of the ODA on travel time.

The second test shows what is achieved with the system in terms of the effects and benefits of implementing such a system when the proposed system is activated and turned off. The test examined the relationship between low, medium, and high levels of traffic and the average travelling time.

The results indicate the mean values for both systems when the system is deployed and turned off. The results in Fig. 5 show that delays are always occurring, particularly when the traffic becomes heavier when the system is off; meanwhile the proposed system can help drivers obtain improved information on the best space they can occupy without surprises.

The third experiment shown in Fig. 6 examines the impact of the ODA on travel time, while more ODA provides further information about the vehicle around the deciding vehicle, thus leading to optimum decisions. The experiment was conducted with a maximum of 50 ODAs; as the moving vehicle loses its connection to the fixed RSU when reaching this number, this is the highest level achieved. However, when using ODA, the decision from a single ODA may not be sufficient, and the vehicle may sometimes send multiple ODAs to make the best decision.

## 6. Conclusions and future work

V2V drivers frequently turn lanes according to visual distance expectations, but many of them are not too well-planned-as the visible range is limited, making it difficult to achieve such a task. This research proposed an assistant lane selection model to help a driver choose an optimal lane travel direction to reduce driving time. The system can be used for the estimation of future conditions of road cell traffic by spatial and temporal details. The simulation results reveal that, depending on the congestion level, the proposed system has good performance scores ranging from 12.5% to 20%. Furthermore, the simulation results indicate that the heavy-traffic scenarios with richer traffic information can lead to optimal decisions.

A further experiment was conducted to compare the effect when the system is deployed and when it is off. As revealed by the results, when the system is off, there is always a delay, whereas system performance is much higher when the system is on. The third experiment shows the impact of the ODA. More ODAs mean more information that can help drivers arrive at optimum decisions. In order to enhance the robustness of the system, the security of the system should be investigated in future works.

### Conflict of interest

The authors declare that they have no known competing financial interests or personal relationships that could have appeared to influence the work reported in this paper.

### Acknowledgments

This research is funded by the Deanship of Research in Zarqa University /Jordan.

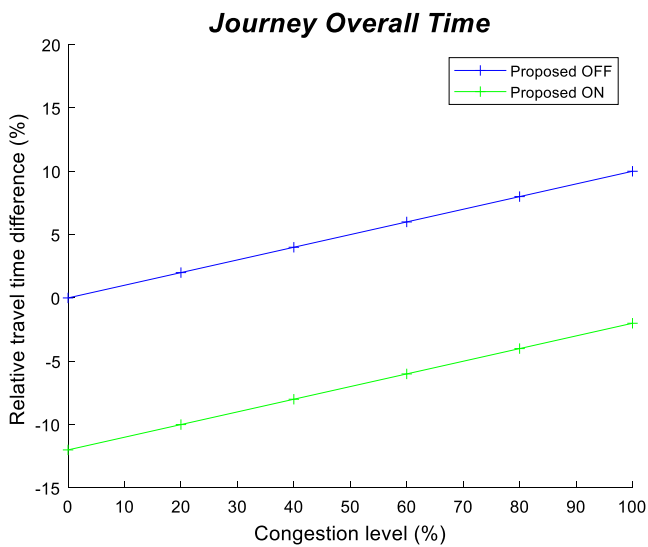

Fig. 5. Proposed system effect.